\documentclass[pdflatex,sn-mathphys-num,12pt]{sn-jnl}

\usepackage{graphicx}%
\usepackage{multirow}%
\usepackage{amsmath,amssymb,amsfonts}%
\usepackage{amsthm}%
\usepackage{mathrsfs}%
\usepackage[title]{appendix}%
\usepackage{xcolor}%
\usepackage{textcomp}%
\usepackage{manyfoot}%
\usepackage{booktabs}%
\usepackage{bold-extra}
\usepackage{caption}
\usepackage{tabularx}

\newcommand{\note}[1]{} 

\begin{document}

\title[From FAIR to CURE: Guidelines for Computational Models of Biological Systems
]{From FAIR to CURE: Guidelines for Computational Models of Biological Systems
}


\author*[1,4]{\fnm{Herbert M.} \sur{Sauro}}\email{hsauro@uw.edu}


\author[10]{\fnm{Eran} \sur{Agmon}} \email{agmon@uchc.edu}

\author[10]{\fnm{Michael L.}  \sur{Blinov}} \email{blinov@uchc.edu}

\author[23]{\fnm{John H.} \sur{Gennari}} \email{gennari@uw.edu}

\author[4]{\fnm{Joe} \sur{Hellerstein}} \email{joseph.hellerstein@gmail.com}

\author[1]{\fnm{Adel} \sur{Heydarabadipour}} \email{adelhp@uw.edu}

\author[14]{\fnm{Peter} \sur{Hunter}} \email{p.hunter@auckland.ac.nz}

\author[1]{\fnm{Bartholomew E.} \sur{Jardine}} \email{barthj@uw.edu}

\author[31]{\fnm{Elebeoba} \sur{May}} \email{emay5@wisc.edu}

\author[14]{\fnm{David P.} \sur{Nickerson}} \email{d.nickerson@auckland.ac.nz}

\author[1]{\fnm{Lucian P.} \sur{Smith}} \email{lpsmith@uw.edu}


\author[24]{\fnm{Gary} D \sur{Bader}} \email{gary.bader@utoronto.ca}

\author[34]{\fnm{Frank} \sur{Bergmann}} \email{frank.bergmann@bioquant.uni-heidelberg.de}

\author[1,3,4,5]{\fnm{Patrick M.} \sur{Boyle}} \email{pmjboyle@uw.edu}

\author[18,19,20]{\fnm{Andreas} \sur{Dräger}} \email{andreas.draeger@informatik.uni-halle.de}

\author[28]{\fnm{James R.} \sur{Faeder}} \email{faeder@pitt.edu}

\author[13]{\fnm{Song} \sur{Feng}} \email{song.feng@pnnl.gov}

\author[22]{\fnm{Juliana}  \sur{Freire}} \email{juliana.freire@nyu.edu}

\author[29]{\fnm{Fabian}  \sur{Fröhlich}} \email{fabian.frohlich@crick.ac.uk}

\author[2]{\fnm{James A.} \sur{Glazier}} \email{jaglazier@gmail.com}

\author[6]{\fnm{Thomas E.} \sur{Gorochowski}}
\email{thomas.gorochowski@bristol.ac.uk}

\author[38]{\fnm{Tomas} \sur{Helikar}}
\email{thelikar2@unl.edu}

\author[35]{\fnm{Stefan} \sur{Hoops}}
\email{shoops@virginia.edu}

\author[1]{\fnm{Princess} \sur{Imoukhuede}} \email{pii@uw.edu}

\author[33]{\fnm{Sarah M.} \sur{Keating}} \email{s.keating@ucl.ac.uk}

\author[32]{\fnm{Matthias} \sur{Konig}} \email{koenigmx@hu-berlin.de}

\author[30]{\fnm{Reinhard}  \sur{Laubenbacher}} \email{reinhard.laubenbacher@medicine.ufl.edu}

\author[10]{\fnm{Leslie M.}  \sur{Loew}} \email{les@uchc.edu}

\author[17]{\fnm{Carlos F.} \sur{Lopez}}\email{clopez@altoslabs.com}

\author[7,8]{\fnm{William W.} \sur{Lytton}} \email{billl@neurosim.downstate.edu}

\author[39]{\fnm{Andrew} \sur{McCulloch}} \email{amcculloch@ucsd.edu}

\author[10]{\fnm{Pedro} \sur{Mendes}} \email{pmendes@uchc.edu}

\author[27]{\fnm{Chris J.}  \sur{Myers}} \email{chris.myers@coloardo.edu}

\author[9]{\fnm{Jerry G.} \sur{Myers}} \email{jerry.g.myers@nasa.gov}

\author[25,26]{\fnm{Lealem}  \sur{Mulugeta}} \email{lealem@insilico-labs.com}

\author[15,16]{\fnm{Anna} \sur{Niarakis}\email{anna.niaraki@univ-tlse3.fr}}

\author[31]{\fnm{David D.} \sur{van Niekerk}} \email{ddvniekerk@sun.ac.za}

\author[11]{\fnm{Brett G.} \sur{Olivier}} \email{b.g.olivier@vu.nl}

\author[10]{\fnm{Alexander A.}  \sur{Patrie}} \email{apatrie@uchc.edu}

\author[2]{\fnm{Ellen M.} \sur{Quardokus}} \email{ellenmq@iu.edu}

\author[40]{\fnm{Nicole} \sur{Radde}} \email{Nicole.Radde@isa.uni-stuttgart.de}

\author[37]{\fnm{Johann M.} \sur{Rohwer}} \email{jr@sun.ac.za}

\author[36]{\fnm{Sven} \sur{Sahle}} \email{sven.sahle@bioquant.uni-heidelberg.de}

\author[10]{\fnm{James C.} \sur{Schaff}} \email{schaff@uchc.edu}

\author[30]{\fnm{T.J.}  \sur{Sego}} \email{timothy.sego@medicine.ufl.edu}

\author[1]{\fnm{Janis} \sur{Shin}} \email{jshin1@uw.edu}

\author[37]{\fnm{Jacky L.} \sur{Snoep}} \email{jls@sun.ac.za}

\author[12]{\fnm{Rajanikanth} \sur{Vadigepalli}} \email{rajanikanth.vadigepalli@jefferson.edu}

\author[13]{\fnm{H. Steve} \sur{Wiley}} \email{Steven.Wiley@pnnl.gov}

\author[41]{\fnm{Dagmar} \sur{Waltemath}} \email{dagmar.waltemath@uni-greifswald.de}

\author[10]{\fnm{Ion} \sur{Moraru}}\email{moraru@uchc.edu}


\affil*[1]{\orgdiv{Department of Bioengineering}, \orgname{University of Washington}, \orgaddress{Seattle}, \postcode{98195-5061}, \state{WA}, \country{USA}}

\affil[2]{\orgdiv{Intelligent Systems Engineering and Biocomplexity Institute}, \orgname{Indiana University}, \orgaddress{\street{Street}, \city{Bloomington}, \postcode{47408}, \state{Indiana}, \country{USA}}}

\affil[3]{\orgdiv{Center for Cardiovascular Biology}, \orgname{University of Washington}, \orgaddress{Seattle}, \postcode{98195-5061}, \state{WA}, \country{USA}}

\affil[4]{\orgdiv{eScience Institute}, \orgname{University of Washington}, \orgaddress{Seattle}, \postcode{98195-5061}, \state{WA}, \country{USA}}

\affil[5]{\orgdiv{Institute for Stem Cell and Regenerative Medicine}, \orgname{University of Washington}, \orgaddress{Seattle}, \postcode{98195-5061}, \state{WA}, \country{USA}}

\affil[6]{\orgdiv{School of Biological Sciences}, \orgname{University of Bristol}, \orgaddress{24 Tyndall Avenue}, \city{Bristol}, \postcode{BS8 1TQ}, \country{UK}}

\affil[7]{\orgdiv{Departments of Physiology \& Pharmacology, Neurology}, \orgname{Downstate Health Science University}, \orgaddress{Brooklyn}, \postcode{11203}, \state{NY}, \country{USA}}

\affil[8]{\orgdiv{Department of Neurology}, \orgname{Kings County Hospital}, \orgaddress{{Brooklyn}, \postcode{11203}, \state{NY}, \country{USA}}}

\affil[9]{\orgdiv{NASA-John H. Glenn Research Center}, \orgaddress{MS 110-3}, \orgaddress{21000 Brookpark Road}, \orgaddress{Cleveland}, \postcode{44135}, \state{Ohio}, \country{USA}}

\affil[10]{\orgdiv{Center for Cell Analysis and Modeling}, \orgname{UConn Health}, \orgaddress{263 Farmington Avenue}, \orgaddress{Farmington}, \postcode{06030-6406}, \state{Connecticut}, \country{USA}}

\affil[11]{\orgdiv{Amsterdam Institute for Life and Environment}, \orgname{Vrije Universiteit Amsterdam}, \orgaddress{De Boelelaan 1108}, \postcode{1081 HZ}, \city{Amsterdam}, \country{Netherlands}}

\affil[12]{\orgdiv{Department of Pathology and Genomic Medicine}, \orgname{Thomas Jefferson University}, \orgaddress{1020 Locust St}, \orgaddress{Philadelphia}, \postcode{19107}, \state{Pennsylvania}, \country{USA}}

\affil[13]{\orgdiv{Biological Sciences Division}, \orgname{Pacific Northwest National Laboratory}, \orgaddress{902 Battelle Blvd}, \orgaddress{Richland}, \postcode{99354}, \state{WA}, \country{USA}}

\affil[14]{\orgdiv{Auckland Bioengineering Institute}, \orgname{University of Auckland}, \orgaddress{Auckland}, \postcode{1010}, \country{New Zealand}}

\affil[15]{\orgdiv{Molecular, Cellular and Developmental Biology Unit (MCD)}, \orgname{Center of Integrative Biology}, \orgname{University of Toulouse III-Paul Sabatier}, \orgaddress{165 Rue Marianne Grunberg-Manago}, \orgaddress{Toulouse}, \postcode{31400}, \country{France}}

\affil[16]{\orgdiv{Lifeware Group}, \orgname{Inria}, \orgaddress{Building Alan Turing}, \orgaddress{1 Rue Honoré d'Estienne d'Orves}, \postcode{91120}, \city{Palaiseau}, \country{France}}

\affil[17]{\orgdiv{Multiscale Modeling Group}, \orgname{Altos Labs}, \postcode{94065}, \city{Redwood City}, \state{CA}, \country{USA}}

\affil[18]{\orgdiv{German Center for Infection Research (DZIF)}, \orgname{partner site Tübingen}, \city{Tübingen}, \country{Germany}}

\affil[19]{\orgdiv{Eberhard Karl University of Tübingen}, \orgname{Quantitative Biology Center (QBiC)}, \orgaddress{Ottfried-Müller-Str. 37}, \postcode{72076}, \city{Tübingen}, \country{Germany}}

\affil[20]{\orgdiv{Martin Luther University Halle-Wittenberg}, \orgname{Data Analytics and Bioinformatics}, \orgaddress{Von-Seckendorff-Platz 1}, \postcode{06120}, \city{Halle (Saale)}, \country{Germany}}

\affil[21]{\orgdiv{Vrije Universiteit Amsterdam}, \orgname{Amsterdam Institute for Life and Environment}, \orgaddress{De Boelelaan 1108}, \postcode{1081 HZ}, \city{Amsterdam}, \country{Netherlands}}

\affil[22]{\orgdiv{Department of Computer Science and Center for Data Science}, \orgname{New York University}, \orgaddress{New York, NY}, \postcode{11201}, \city{New York}, \country{USA}}

\affil[23]{\orgdiv{Department of Biomedical Informatics \& Medical Education}, \orgname{University of Washington}, \orgaddress{1959 NE Pacific Street}, \postcode{98195}, \city{Seattle}, \state{Washington}, \country{USA}}

\affil[24]{\orgdiv{The Donnelly Centre}, \orgname{University of Toronto}, \orgaddress{160 College St}, \postcode{M5S 3E1}, \city{Toronto}, \state{Ontario}, \country{Canada}}

\affil[25]{\orgdiv{InSilico Labs LLC}, \orgname{InSilico Labs LLC}, \postcode{77008}, \city{Houston}, \state{Texas}, \country{USA}}

\affil[26]{\orgdiv{Medalist Performance}, \postcode{77027}, \city{Houston}, \state{Texas}, \country{USA}}

\affil[27]{\orgdiv{Department of Electrical, Computer, and Energy Engineering}, \orgname{University of Colorado Boulder}, \orgaddress{425 UCB}, \orgaddress{Boulder}, \postcode{80309}, \state{Colorado}, \country{USA}}

\affil[28]{\orgdiv{Department of Computational and Systems Biology}, \orgname{University of Pittsburgh}, \orgaddress{3500 Fifth Avenue}, \postcode{15213}, \city{Pittsburgh}, \state{Pennsylvania}, \country{USA}}

\affil[29]{\orgdiv{Dynamics of Living Systems Laboratory}, \orgname{The Francis Crick Institute}, \orgaddress{1 Midland Road}, \postcode{NW1 1AT}, \city{London}, \country{UK}}

\affil[30]{\orgdiv{Department of Medicine}, \orgname{University of Florida}, \orgaddress{1600 SW Archer Rd}, \postcode{32610-0225}, \city{Gainesville}, \state{Florida}, \country{USA}}

\affil[31]{\orgdiv{Wisconsin Institute for Discovery}, \orgname{University of Wisconsin—Madison}, \orgaddress{330 North Orchard Street}, \postcode{53715}, \city{Madison}, \state{WI}, \country{USA}}

\affil[32]{\orgdiv{Institute for Biology, Institute for Theoretical Biology}, \orgname{Humboldt-University Berlin}, \orgaddress{Philippstraße 13}, \postcode{10115}, \city{Berlin}, \country{Germany}}

\affil[33]{\orgdiv{Advanced Research Computing Centre}, \orgname{University College London}, \orgaddress{Philippstraße 13}, \postcode{WC1E 6BT}, \city{London}, \country{UK}}

\affil[34]{\orgdiv{COS Heidelberg}, \orgname{Heidelberg University}, \orgaddress{Im Neuenheimer Feld 230}, \postcode{69120}, \city{Heidelberg}, \country{Germany}}

\affil[35]{\orgdiv{Biocomplexity Institute}, \orgname{University of Virginia}, \orgaddress{Town Center Four, 3rd Floor}, \orgaddress{994 Research Park Boulevard}, \postcode{22911}, \city{Charlottesville}, \state{VA}, \country{USA}}

\affil[36]{\orgdiv{BioQuant}, \orgaddress{Im Neuenheimer Feld 267}, \postcode{69120}, \city{Heidelberg}, \country{Germany}}

\affil[37]{\orgdiv{Department of Biochemistry}, \orgname{University of Stellenbosch}, \orgaddress{Private Bag X1}, \postcode{7602}, \city{Matieland}, \country{South Africa}}

\affil[38]{\orgdiv{Department of Biochemistry}, \orgname{University of Nebraska-Lincoln}, \orgaddress{Beadle Center},  \postcode{68588-0664}, \city{Lincoln NE}, \country{USA}}

\affil[39]{\orgdiv{Departments of Bioengineering and Medicine}, \orgname{University of California San Diego}, \orgaddress{9500 Gilman Drive}, \postcode{92093-0412}, \city{La Jolla}, \state{CA}, \country{USA}}

\affil[40]{\orgdiv{Institute for Stochastics and Applications}, \orgname{University of Stuttgart}, \orgaddress{Pfaffenwaldring 9}, \postcode{70569}, \city{Stuttgart}, \country{Germany}}

\affil[41]{\orgdiv{Medical Informatics Laboratory}, \orgname{University Medicine Greifswald}, \orgaddress{D-17489}, \city{Greifswald}, \country{Germany}}

\abstract{Guidelines for managing scientific data have been established under the FAIR principles requiring that data be Findable, Accessible, Interoperable, and Reusable. In many scientific disciplines, especially computational biology, both data and {\em models} are key to progress. For this reason, and recognizing that such models are a very special type of ``data'', we argue that computational models, especially mechanistic models prevalent in medicine, physiology and systems biology, deserve a complementary set of guidelines. We propose the CURE principles, emphasizing that models should be Credible, Understandable, Reproducible, and Extensible. We delve into each principle, discussing verification, validation, and uncertainty quantification for model credibility; the clarity of model descriptions and annotations for understandability; adherence to standards and open science practices for reproducibility; and the use of open standards and modular code for extensibility and reuse. We outline recommended and baseline requirements for each aspect of CURE, aiming to enhance the impact and trustworthiness of computational models, particularly in biomedical applications where credibility is paramount. Our perspective underscores the need for a more disciplined approach to modeling, aligning with emerging trends such as Digital Twins and emphasizing the importance of data and modeling standards for interoperability and reuse. Finally, we emphasize that given the non-trivial effort required to implement the guidelines, the community moves to automate as many of the guidelines as possible.}




\maketitle

\section{Introduction}\label{sec1}

Wilkinson et al. in 2016~\citep{wilkinson2016fair} made a good case (dare we say a fair case) for establishing guidelines for the management of scientific data. They arrived at four guiding principles enshrined in the acronym FAIR, namely that data be Findable, Accessible, Interoperable, and Reusable. With the rapid growth of computational modeling, especially the development of mechanistic models of physiological and cellular systems, the question arises of how these principles can be extended so that they can succinctly describe best practices for model management (e.g., model development, model selection, and model interpretation). In this perspective, we introduce a set of complementary guidelines to FAIR that address the specific needs for mechanistic models. We identify four principles: Credibility, Understandability, Reproducibility, and Extensibility. We refer to these as the \texttt{\textbf{CURE}} principles.

We focus on mechanistic models, thereby excluding the growing body of machine-learned (ML) and AI models that are based solely on data. The machine learning community has appropriately encouraged the use of FAIR principles when publishing ML models~\citep{ravi2022fair}, with an emphasis on ensuring that data are accessible to those who wish to repeat the study. However, the reproducibility of those models is a separate topic that has its own special concerns (e.g., the selection of training, validation, and test data, as well as the choice of hyperparameters). We note in passing that although ML models rarely consider mechanisms, there are situations in which mechanistic models make use of machine learning approaches, such as in the context of parameter estimation or physically informed neural networks (PINNs)~\cite{raissi2019physics}.

Although the focus of this proposal is on models from the systems biology community, the guidelines and sentiments we describe are broadly applicable to other modeling domains.

\section{Why have Guidelines?}\label{sec2}

Models are indispensable in many science and engineering disciplines. Examples include circuit simulation in electrical engineering, models of fluid flow in mechanical engineering, and weather modeling in atmospheric science. In some cases, modeling has progressed to the development of digital twins~\citep{juarez2021digital,laubenbacher2024toward}, in which a simulation model is designed to replicate and interact with the physical system, or virtual populations, which can be used for applications such as clinical trial design~\cite{Michael2023-kb}. Biological modeling has not yet risen to that same level of sophistication, with the possible exception of areas such as protein folding~\citep{dill2012protein} and molecular dynamics~\cite{mccammon1977dynamics} where physics and chemistry play a more important role. Even so, biological modeling is a rapidly evolving field. As the field grows, guidelines in the spirit of FAIR will help modelers create more impactful and credible models. We believe these guidelines will be of importance for models that ultimately reach the clinic and particularly for the growing interest in developing biomedical Digital Twins. The Working Group `Building Immune Digital Twins' tries to address these questions, having as a focus the human immune system and its responses in various pathological contexts~\citep{Niarakis2024, Laubenbacher2022}. 


\section{Existing Guidelines}\label{sec4}


Several groups have proposed guidelines to improve best practices in creating biological models over the last 20 years. Of particular note is the creation of standardised languages for biological models such as Systems Biology Markup Language (SBML)~\citep{hucka2003systems}, CellML~\citep{clerx2020cellml} and NeuroML~\citep{gleeson2010neuroml}. These are machine-readable formats that are an {\em explicit representation} of the model. By explicit representation, we mean that the model representation only includes elements that are essential for modeling; it does not include implementation details related to simulation. For example, the essential characteristics of a mechanistic model of a well-mixed biochemical system include chemical species, reactions, and rate laws. It does not include software details such as file input/output and control logic for numerical solvers. An explicit representation is {\em independent of its implementation} in software.

The choice of an explicit representation for models is driven by the requirements of the communities that develop and use the models. SBML focuses on biochemical models where the representation is in terms of the biological processes. CellML focuses on a mathematical representation of models as differential-algebraic equations. NeuroML focuses on representations of neural systems. These representations have become popular among modelers and software developers. For example, all genome-scale models~\citep{gu2019current} are represented using SBML, and thousands of kinetic models are now stored on publicly accessible model repositories using these formats~\citep{king2016bigg,arkin2018kbase}. Standards such as SBML avoid the use of potentially confusing and unreusable {\em ad hoc} models, allowing models to persist in a reproducible form over long periods of time~\citep{blinov2021practical,porubsky2021publishing}. Many authors, however, still publish models in executable formats such as MATLAB, Python, etc., which can pose problems for reproducibility and reuse, particularly when poorly documented~\citep{tiwari2021reproducibility,erdem2022scalable}. The logical modeling community, which uses SBML Qualitative format (SBML-Qual) to encode logical and Boolean models, made efforts to define a roadmap toward the annotation and curation of logical models (aka the CALM initiative), including milestones for best practices and recommended standard requirements ~\citep{10.1093/bib/bbaa046}.

To promote data sharing and reuse, the FAIR principles recommend a data dictionary that specifies data types and semantics for data items~\citep{wilkinson2016fair}. An analogous requirement exists for models. For example, consider an SBML model with the reaction $A \rightarrow B$. Annotations are used to define $A$, $B$, and provide information about the reaction (e.g., the organism, cell type, and organelle in which the reaction takes place).
Annotation can provide additional ontological and reference information about a model, including its submodels, processes, assumptions, and provenance. Genome-scale models are heavily annotated with process metadata~\cite{blais2013linking,passi2021genome}, and the curators at BioModels~\citep{malik2020biomodels} regularly add annotations to curated models. As part of these efforts, the systems biology and physiology community developed the MIRIAM standard, which describes the ``Minimum Information Required In the Annotation of Models''~\cite{novere2005minimum}. MIRIAM applies to structured information, such as SBML or CellML, where annotation information can be included in a machine-readable manner. Such information can be of great utility for searching, merging, or disassembling a model into its component parts~\citep{neal2014ac}.

\section{Mechanistic Models}\label{sec6}

The focus of this perspective is on mechanistic models\note{it is worth defining what we mean by a mechanistic model. This is more important today given the increased interest in machine learning and AI where mechanism no longer takes center stage, if at all.}. We define a mechanistic model as: a representation of a biological system that is described in terms of the constituent physical parts and processes that occur between the parts. For example, in a mechanistic model of a cell signaling pathway, we would specify the various proteins and their phosphorylation states and the transformations between these states via enzymatic processes involving kinases and phosphatases. 

Often, a mechanistic model is transformed into a computational model by invoking physical laws, such as the conservation of mass and chemical kinetic laws that govern individual transformations. In physiology and systems biology, models are commonly represented as a system of ordinary differential equations~\citep{ingalls2013mathematical,sauro2020systems}, but other representations are also used, such as systems of Boolean functions, graphs, stochastic systems, and constraint-based models~\citep{kim2018review}. Such models are often shared via model repositories such as BioModels~\citep{malik2020biomodels}, JWS Online~\cite{peters2017jws}, or BiGG~\citep{king2016bigg} and KBase~\cite{arkin2018kbase} for constraint-based models. 

Models can also be shared using raw executable formats such as Python or MATLAB that describe differential equations. Such `raw' model code is typically provided as supplementary files to a published paper or stored in code repositories such as GitHub or specialized repositories such as ModelDB~\citep{mcdougal2017twenty}. In some cases, the model may be described as part of the main text of the published article~\citep{choi2011physiological} or be absent altogether. This obviously makes the reproducibility of such work much more complex and sometimes impossible, given the frequency of typological errors in printing mathematical equations or code fragments. Models-as-programs paradigms such as those followed by PySB and others encourage the use of best practices in Python coding and documentation through tools such as Sphinx, but these approaches rely on developer effort to document the code and model at an appropriate level of detail~\citep{lopez2013programming, brandl2021sphinx}. They also tie a model to a specific language implementation, making reuse difficult and backward compatibility a problem.

It may be difficult to abstract the underlying physiological processes of some complex systems purely in explicit form, though there have been few efforts to attempt this.  
Multi-scale modeling tends to give rise to such challenging scenarios.
The context of computational modeling of electrophysiological phenomena in the heart provides an illustrative example~\cite{plank2021opencarp}. Ordinary differential equations describing cell-scale events like ion channel gating and the generation of action potentials can be encoded using CellML or as a biological description in SBML. However, representing the propagation of excitatory wavefronts requires spatial discretization of the governing partial differential equations via finite element analysis; explicit formats for such applications exist~\cite{britten2013fieldml,schaff2023sbml} but have not seen widespread use.  One approach to improving inter-operability and reproducibility is to create tools for importing model components written in common data formats.
For instance, the openCARP simulation environment~\cite{plank2021opencarp} includes a CellML ``translator'' that facilitates on-the-fly incorporation of cell-scale representations of different types of cardiac electrophyiology (e.g., cardiac region, species, or disease-specific action potentials) within the fabric of the multi-scale simulation ecosystem.

Mechanistic models of biological systems often contain many parameters whose values are unknown (e.g., kinetic constants and diffusion rates). A popular approach to estimating these parameters is model fitting or calibration~\citep{mendes1998non,gabor2015robust,ashyraliyev2009systems,sauro2020systems}. This is where model parameters are adjusted, often with an optimization algorithm, so that the output of the model matches relevant experimental data. Because models tend to have many parameters, the problem is often, if not always, underdetermined. This means that the set of fitted parameters in a model is not unique. In some instances, one or more parameters may be non-identifiable, meaning no single value can be assigned to the parameter~\citep{wieland2021structural}. Solutions to this problem can involve simplifying the model to reduce the number of parameters and/or eliminate non-identifiable parameters; another approach is to collect additional experimental data. More recent methods have focused on Bayesian techniques to assess the uncertainty in parameter estimates~\citep{shockley2018pydream,irvin2023model}. 

As a side note, we contrast mechanistic models with the great variety of neural network models that can provide accurate predictions for complex problems~\citep{nikolados2022}. However, the resulting ``black box'' models are almost always challenging to understand, a situation referred to as ``intellectual debt''~\citep{zittrain2019}. An example is the prediction of COVID-19 from chest x-rays~\cite{islam2021accurate}. A close examination of the model revealed that it ignored the medical features in the x-rays and instead relied on the coding of the hospital at which the imaging was done. This turned out to be a surrogate variable for the patient population since one hospital had more COVID patients than the other. We cite this as an example of a model with accurate predictions that was not credible.

\section{The CURE Guidelines}\label{sec8}

As with FAIR, we define four specific guidelines to improve best practices in developing mechanistic computational models of biological systems. There is no specific order in the guidelines but the acronym CURE seemed appropriate for the topic in question.

\begin{figure}[h]
\centering
\includegraphics[scale=0.5]{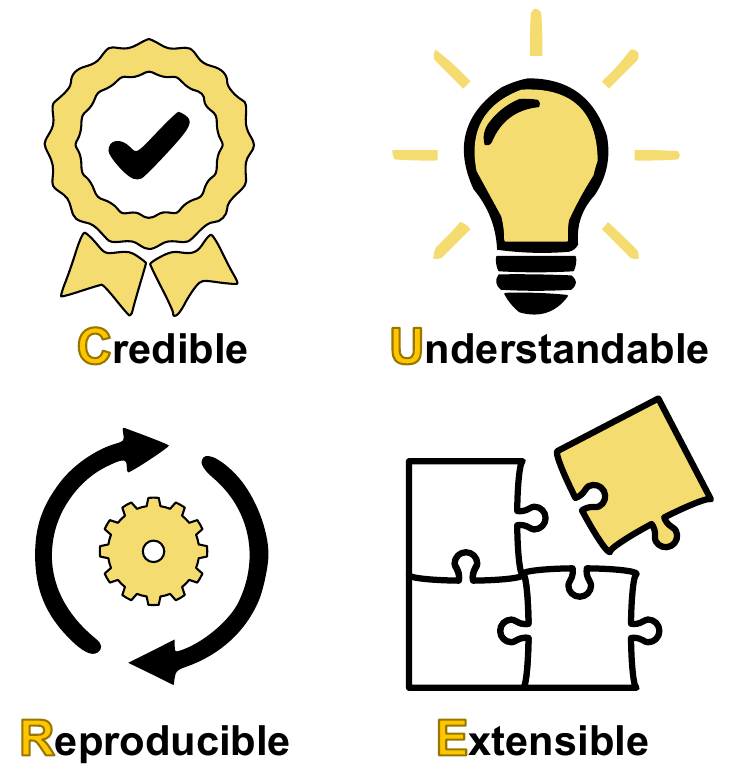}
\end{figure}

CURE\footnote{Another organization devoted to the promotion of curation practices of research compendia also uses the acronym CURE:``Curating for Reproducibility''. However, there is little overlap between the two usages.} covers four key ideas, \textbf{Credibility}, \textbf{Understandability}, \textbf{Reproducibility}, and \textbf{Extensibility}, which we describe in the following sections. These guidelines are meant, where possible, to apply to both machine-readable formats, such as SBML, as well as models distributed in executable code such as MATLAB or Python. They also align with previous community efforts to address barriers in comprehensiveness, accessibility, reusability, interoperability, and reproducibility of computational models in systems biology~\citep{10.1093/bib/bbac212}. 

\subsection{Credible} 

We use credibility to mean a perceived measure of believability~\citep{yilmaz2022model,tatka2023adapting}. A credible model makes trustworthy and actionable predictions within the range of conditions it was intended to simulate. Prior work on model credibility dates back to 1979~\citep{schlesinger1979terminology}, when the Society for Modeling \& Simulation International (SCS) described many concepts associated with model credibility. More recently, several groups within the biological modeling community have discussed model credibility, most notably the ten rules developed by CPMS~\citep{erdemir2020credible}, devised by the Committee on Credible Practice of Modeling \& Simulation in Healthcare (CMPS) in the US, and in Europe Musuamba et al~\citep{musuamba2021scientific} who describe in some detail criteria and concepts that are important to assess the model credibility. Of note, the CPMS working group considered credibility as a descriptor of the practice of modeling and simulation rather than that of a model. Accordingly, the assessment by the ten CPMS rules are geared towards evaluating the practices followed in the modeling efforts~\citep{vadigepalli2024rubric}. Two core concepts related to model credibility are verification and validation~\citep{thacker2004concepts}.

\textbf{Verification} is ``the process of determining that a model implementation accurately represents the developer’s conceptual description of the model and the solution to the model''~\citep{thacker2004concepts}. Similar definitions can be found in many other documents~\citep{national2012assessing}. In practice, verification means assessing the correctness of: (i) the model representation (e.g., detecting typographical errors); (ii) the numerical algorithms used to simulate the model (e.g., pseudo random numbers have the correct distributions); and (iii) the model implementation in software. Models implemented directly in programming languages such as Python may not have (i), and models implemented using standards such as SBML have few concerns about (iii).

When using standards such as SBML, verification involves ensuring that different software applications interpret the description of the model in the same way~\citep{bergmann2008comparing} and that the software has passed the SBML test suite~\citep{sbmltestsite}. It is worth noting the close correspondence of verification to software testing, including unit and system tests, and documentation~\citep{hellerstein2019recent}. Since models are almost always implemented in software, this should come as no surprise. Static tests can assess whether the model is correct, for example, that the biophysical laws have been entered correctly or that mass balance is not violated. Dynamic tests can offer more subtle checks to the correctness of a model. For example, a model whose variables are concentrations should not be able to reach negative values. 

\textbf{Validation} is ``the process of determining the degree to which a model is an accurate representation of the real world from the perspective of the intended uses of the model''~\citep{thacker2004concepts}. Of significant importance is the phrase ``intended uses''. All models have a limited scope (including AI/ML models). That is, they can only make useful predictions within their intended purpose and calibration. This is particularly important for models that might be used in a clinical setting where misuse of a model could have dire consequences. It is important, therefore, that there be a clear statement of the purpose of the model as well as the conditions under which the model is applicable. Validation involves comparing experimental data to predictions made by the model. In practice, given the nature of scientific models, not all model predictions can be validated. Validation is more a measure of confidence in a model's credibility to match reality than an absolute statement of truth. 

As noted previously, mechanistic models often have parameters whose values are determined through parameter fitting The credibility of the parameter estimates can impact the credibility of the model 
and can be enhanced in two ways, either by cross-validation or through the use of competing models. During cross-validation, the experimental dataset is split into a training and a test set. The test set is used to check whether the calibrated model can recapitulate the test set. If the model can recover the test set then there is greater confidence in the model. If the model fails to recover the test set, then the model is inadequate and needs to be reexamined. Likewise, in the context of multi-scale simulations, model credibility can be assessed by experimentally documenting the system-wide response to at least two perturbations (e.g., stimulating the heart from distinct sites~\cite{corrado2018work}).
Then, a model parameterized using exclusively data based on the first set of measurements can be convincingly validated by demonstrating its ability to reproduce the second set (despite the model never having seen those data during calibration).

For parameter-free logic-based models in biology, the credibility lies in the causality of the statements used to build the logical rules and functions ~\cite{10.1093/bib/bbaa390}, as well as the binarized interpretation and use of small- and larger-scale experimental data ~\cite{HALL2021100386}. 

This leads to the question of model selection~\citep{kirk2013model}. That is, given the limited amount of experimental data, a given model will not necessarily be unique, and other models could equally likely to explain the data. The literature on model selection is extensive, but certain popular tests have emerged, most notably the Akaike information criterion (AIC)~\citep{akaike1974} and the Bayes Information Criterion (BIC)~\citep{beik2023unified}. The AIC is an approximation, based on maximum likelihood, to the Kullback-Leibler divergence~\citep{kullback1951}, which quantifies the difference between the model and full reality. The AIC test considers both the quality of a fit and the number of parameters in the model. For example, given two models that fit the data equally well within experimental error, the model with fewer parameters is the preferred. Any number of plausible models can be compared in this way since the computation is relatively straightforward. Kirk et al.~\citeauthor{kirk2013model} provide an excellent review of this topic~\citep{kirk2013model}. Model credibility can be enhanced if, given the uncertainty in the underlying biology, a number of models are proposed, with a measure such as AIC, used to indicate their relative credibility. The BIC is based on the relationship to the Bayes Factor, itself the ratio of the likelihood of two hypotheses and used to compare the marginal likelihoods between two models. 



\textbf{Uncertainty Quantification.} 
In recent years, there has been a growing interest in the biomedical community to use uncertainty quantification (UQ) as part of a credibility assessment~\citep{viceconti2021silico}. UQ has been well established in other scientific domains for many decades~\citep{booker2011evolution} and involves calculating how uncertainty in the experimental data and model parameters contributes to uncertainty in the model outputs. A model that generates highly uncertain outputs is seen as less credible. Verification, validation, and UQ have been collectively referred to by many practitioners using the acronym VVUQ~\citep{national2012assessing}. A recent article by Colebank et al., reviews elements of this topic~\ref{colebank2024guidelines}.

Another criterion that can enhance a model's credibility is the provenance of the data used to build the model. Where did the data come from, and was it modified in some way? Table~\ref{tbl:credibility} gives a summary of some of the main criteria that can be used to access credibility. However, not all the criteria are equally weighted. Validation and verification are often the most important in this regard. The first four rules in the 10 rules devised by the Committee on Credible Practice of Modeling \& Simulation emphasize all these aspects~\citep{erdemir2020credible}.  

\begin{table}[htpb]
\begin{tabular}{lp{10cm}} \toprule
Attribute & Criteria \\ \midrule
Validation & How well do the model outputs match reality? \\ 
Verification & Has the model been constructed without error; are the simulation algorithms correctly encoded and operating without error? \\
Uncertainty & Have the effects of uncertainty in the model outputs been reported? \\
Provenance & Can the data that was used to calibrate or validate the model be traced to its original source? \\
Annotation & Have the inputs and output of the model been well defined? \\
Assumptions & Have the assumptions used in the model been made explicit? \\
Purpose & Has the purpose of the model been adequately described? \\
Scope & Has the scope, that is, the space within which the model can be used, been specified? \\
Unbiased & Was the model calibrated on unbiased data? This depends on the scope of the model, but if a model is to be used across a diverse population, then clearly, the calibration data needs to be diverse. \\ \bottomrule
\end{tabular}
\caption{A range of criteria that can establish the credibility of a model.}
\label{tbl:credibility}
\end{table}

In many cases, modelers may opt not to consider some or all of these criteria in their work, primarily due to the burden of having to do the checks. We, therefore, recommend automating as many of these criteria as possible. For example, verification of SBML-based models can be achieved using the BioSimulation resource~\citep{shaikh2022biosimulators}. Validation tests could be provided in a standard format as part of the software modeling code just as software engineers often provide validation tests as part of their distributions~\citep{pythonUnitTests}. More difficult is including information about data provenance and model assumptions. However, the use of model standards such as SBML or CellML do allow models to be annotated with this information. The same applies to indicating the scope and purpose of a model. When models have an explicit representation (e.g., SBML) as opposed to just a software implementation, analysis tools can do deep dives into the model to examine its biophysical assumptions. Also, more in-depth verification can be done on the explicit model representation, such as detecting errors in the formulation of biophysical laws~\citep{hoops2006copasi}. MEMOTE~\citep{lieven2020memote} is a successful example of automated software that can do deep dives into genome-scale models to assess the quality of the model.

\note{Is it worth making the distinction between the credibility of the model -- VVUQ, and the credibility of the process that created the model -- the focus of most of the 10 CPMS rules. It is certainly possible, even easy, to do everything right in terms of process but end up with a model with no validity.}
Credibility is widely used in software development through the commenting of code, using version control for provenance, and unit and system testing for validation~\citep{hellerstein2019recent}. Verification is achieved through in-depth checking of the software compilers and runtime systems.

\subsection{Understandable}
One of the aims of the scientific method is to gain an understanding of how the world works by proposing models and theories to be tested. We understand theories (to paraphrase~\citep{fried2020theories}) to be bodies of knowledge that are broad in scope. Chemical reaction theory or the central dogma in biology are examples of broad scope. In contrast, models are instantiations of theories and, as a result, are narrower in scope and often represent a particular biological process of interest. This includes computational models of metabolism, protein signaling networks, etc. Both models and theories are the most important outcome of the scientific method. They provide a way to rationalize a set of observations and make predictions about the future state of the system. However, the act of ``understanding''~\citep{de2017understanding} a model or theory is not an easy concept to define and may encompass a number of different aspects. In particular, how might one quantify ``understanding''? Philosophically, de Regt and Dieks~\citep{de2017understanding} define understanding as a given phenomenon if there exists a theory that describes the phenomenon. However, such definitions are hard to quantify. Instead, we will focus on measures that can be used to provide some level of confidence in how understandable a model is. 

We note in passing that understanding modern AI~\citep{chang2024survey,thirunavukarasu2023large} and machine learning algorithms can be very challenging, if not impossible to understand. It may not even be the goal of such approaches where the underlying physical reality is entirely superfluous to the objective. As AI approaches become increasingly accurate at making predictions~\cite{weatherAI2024}, the necessity of understanding how systems work may become entirely unnecessary. There are a number of counterarguments to this view. One is that human beings have an innate desire to understand the world, which is difficult to suppress. In addition, understanding a model can lead to its improvement, which is often transferable knowledge that can be applied to other problems. In a clinical situation, there is no guarantee that training has been sufficient to provide reliable diagnosis and treatment. If the system makes a mistake, we cannot easily determine why. In software engineering, writing understandable code is crucial to minimize maintenance costs, reduce bugs, and make the software more reliable and extensible~\citep{arvanitou2021software,cain2023code,GoodCodePractices}. The obvious counter argument is that the underlying mechanism of many clinical treatments are either poorly understood nor not understood at all and yet they are useful and benefit a great many patients. 



Biological systems, even small ones are challenging to understand due to the nonlinear interactions among the components. The problem gets even more acute as the systems we study get larger. Biological systems often interact in complicated and nonlinear ways that result in emergent behaviors~\cite{nicolis1989exploring}. For example, there is no amount of understanding of the components of DNA -- purines and pyrimidines, or at a lower-scale carbon and hydrogen atoms -- that would lead us to predict its complex structure or understand its biological role; the role emerges from the way that the components are put together.

\begin{figure}
    \centering
    \includegraphics[scale=0.7]{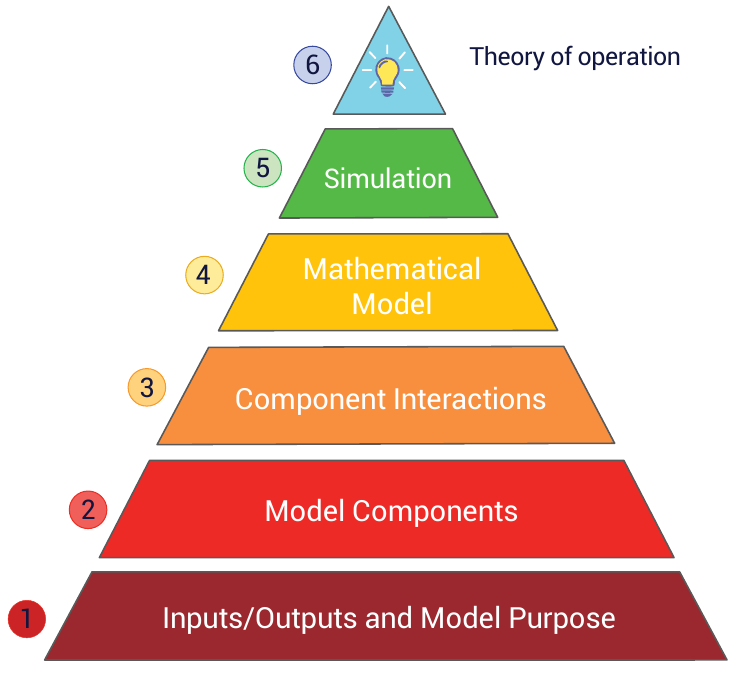}
    \caption{Quantifying understanding through a hierarchy of criteria.}
    \label{fig:pyramid}
\end{figure}

How can we measure understanding? One way is to divide a model's components and attributes into levels of perceived importance. For example, we could understand different aspects of a model such as knowing a model’s purpose, its components, the biophysical laws that describe how the components interact, its inputs, outputs, assumptions, and limitations. Figure~\ref{fig:pyramid}, depicts a pyramid that organizes such a hierarchy of `understanding'. At the base of the pyramid is the most rudimentary understanding; successive layers indicate increased levels of understanding. At the most rudimentary level (1), we want to know the purpose and objectives of the model as well as its inputs and outputs. Subsequent levels include the system components being modeled (2); the interactions between components that are modeled (3); a mathematical description of these interactions (4); a way to evaluate the mathematical model (e.g., solve a system of differential equations) (5); and finally, if possible, a general theory that explains the behavior of the model (6). 

Technology is already available to assist in identifying model components (Level 2) and interactions (Level 3), as well as adding metadata to a model (Level 1). Such information can be provided through model annotations~\cite{le2006model}. The mathematical model (Level 4) also includes the model's assumptions, which can be added as annotations using the SBO ontology~\cite{courtot2011controlled}.\note{I don't see anywhere where 'SBO' or 'KiSAO' are defined; btw, do you want to include a list of acronym on p 1?}
For genome-scale metabolic models, the scientific community developed a list of detailed recommendations to annotate such models at Levels 1 to 4 following an extensive debate at their 2018 annual conference on constraint-based reconstruction and analysis (COBRA)~\citep{Carey2020}.
Scientific models do not have to be mathematical, but for our purposes, most models are, and once a model is defined mathematically, it is often possible to convert the model into an executable form so that simulations can be carried out to generate predictions. Simulation information can be annotated using ontologies such as KiSAO~\cite{zhukova2011kinetic}. The pinnacle of understanding is Level 6, which is the formulation of a theory that explains the behavior seen in the model.
\note{Any interest in mentioning that biological models and theories are not falsifiable, unlike classical popperian physics models? eg DNA makes RNA makes protein is not falsified by retrovirus -- retrovirus is an exception to the theory}

Level 6 is the most difficult to quantify and involves explaining in terms of both fundamental and higher-order concepts how a given behavior emerges. Making this more difficult is that many biological systems show emergent behavior~\citep{nicolis1989exploring}, where we observe behavior that is not found in any individual component of the system. Even simple systems can reach this threshold. For example, a system of only two components and some non-linearity can show sustained energy dependent oscillations~\citep{kholodenko2006cell,sauro2020systems}. Understanding the individual components is not sufficient to explain the origin of the oscillations~\citep{reijenga2005yeast} but requires a theory to help understand how, as a system, we observe oscillations. In this particular case, we need additional concepts, such as hysteresis and negative feedback, to understand the origin of the oscillations. For very complex systems, there may be multiple levels of abstraction that describe many levels of emergent behavior. Such abstractions are commonly used in electrical engineering and computer science. This is what makes it possible to engineer these highly complex systems. Reverse engineering abstraction layers in biology are, however, difficult~\cite{csete2002reverse,andrews2024design} due to the fact we are dealing with evolved systems that are not always as well structured as our own engineered systems. Unfortunately, no ontology exists to adequately annotate how a system operates. TEDDY is closest to an ontology~\cite{courtot2011controlled}, which can be used to describe dynamics but does not, as yet, have the capacity to describe a theory of operation.

\subsection{Reproducible}

Reproducibility is a cornerstone of the scientific method, and over the last 10-15 years, much discussion has been devoted to this topic~\citep{baker20161,plesser2018reproducibility}; mostly about the {\em lack} of reproducibility of many scientific results. Often, we think of wet lab experiments in biology as being difficult to reproducible because they are multifaceted and inherently variable. However, it has also been discovered that computational experiments are often not reproducible~\citep{tiwari2021reproducibility}. This is surprising given that computation involves well-defined and often deterministic procedures. We will not revisit the many issues and recommendations concerning the reproducibility of computational experiments ~\citep{porubsky2020best,porubsky2021publishing,blinov2021practical,shin2023standards,tatka2023adapting} but one thing that has become clear is that community standards such as SBML have significantly improved the reproducibility of computational models by providing a machine-readable representation in a standard format~\citep{mendes2018reproducible,tiwari2021reproducibility}. Moreover, recent evidence supports the opinion that reproducible models are more cited and more likely to be reused in subsequent studies~\citep{hopfl2023bayesian}. Even if reproducibility is not a high priority for its own sake, verification, validation, and reuse of models require a model to be reproducible~\citep{coveney2021reliability}. In recent years we have seen the emergence of software tools and standards that offer support to assist modelers in creating reproducible models~\cite{hoops2006copasi,moraru2008virtual,choi2018tellurium,medley2018tellurium}. One could even go as far as to say that in systems biology at least, the problem of reproducibility is solved~\cite{porubsky2020best}.

\subsection{Extensible and reusable}

Science builds on past efforts, standing on the proverbial shoulders of giants. This is no different when building computational models where past computational models can be enhanced, reused, and further validation applied. Unfortunately, many published models cannot be easily reused~\citep{erdemir2020credible}. This is because many models were published without an explicit representation of the model. Instead, the model is embedded in a software implementation such as MATLAB or Python. The software implementation adds many complexities (e.g., file interfaces and control logic for computing a solution), often in the form of multiple files with minimal documentation. Moreover, models deployed in this way are in a mathematical representation that loses considerable biological information. For example, when a biochemical pathway model is reduced to a set of differential equations the stoichiometric structure of the network is either lost or is difficult to reverse engineer. A further concern is that the mathematical representation greatly complicates the ability to query models to discover similar pathways, kinetics, and other characteristics. These considerations are some of the reasons why genome-scale models are published using SBML~\citep{hucka2003systems,lieven2020memote} so as to preserve as much biological information as possible. 

Hence, the primary concern with publishing a model solely as its software implementation is that it is difficult to reuse the model, either in part or in whole. For example, combining a model written in MATLAB with a model written in Python can be a costly exercise. Likewise, converting models from one format to another, for example, MATLAB to SBML, can also be error-prone and costly~\citep{erdem2022scalable}. Standards such as SBML allow the automated deconstruction~\cite{neal2014ac} and reuse of models~\cite{neal2014reappraisal} through the use of model annotations. Models expressed in SBML are much easier to reuse or extend. When converted to a human-readable language like Antimony~\citep{smith2009antimony}, reuse becomes even easier. 

Other disciplines employ formats such as Modelica~\citep{mattsson1998physical} or representations such as Simulink~\citep{dabney2004mastering} to assist in reuse, but such techniques have not been widely used in developing biological models. 

If executable code is used to represent models, then the model should ideally be partitioned into reusable program functions with ample documentation to illustrate how to reuse the model and what the various symbols in the mathematical equations represent. 

\section{Recommended Requirements}\label{sec11}

The previous discussion provides a wide range of criteria that can be used to satisfy the {\tt CURE} guidelines, and fulfilling all the requirements would be quite onerous. For most academic studies, it might be sufficient to meet a small number of criteria. For models that might be used in a clinical setting, it would be prudent to satisfy as many criteria as possible. Organizations such as the FDA (Food and Drug Administration) have already begun to issue guidelines for models used in medical devices~\citep{food2021assessing}, and there is no reason to doubt that such guidance will eventually be offered for more general use of models in clinical settings. 

For academic research, we therefore wish to propose a recommended set of requirements from each aspect of {\tt CURE} that would be sufficient to significantly impact computational modeling. We provide a checklist in Figure~\ref{fig:checklist}, which also highlights the baseline requirements. 

A key requirement is the need to develop standardized approaches to assess and communicate the extent to which a given model, or a modeling study, satisfies the recommended or baseline requirements. An example to consider is the approach taken by the CPMS working group to develop a rubric that considers the extent of outreach to various stakeholders in satisfying various credible practice guidelines~\cite{vadigepalli2024rubric}. Such evaluations, typically conducted as self-assessments by the respective study authors, can be shared with the community as part of supplementary material in published studies~\cite{verma2018causality,gee2023closed}. Automation could greatly facilitate this assessment process, especially if provided in advance and used during model development. 

\subsection{Baseline Requirements}

\begin{figure}[h]
\includegraphics[scale=0.55]{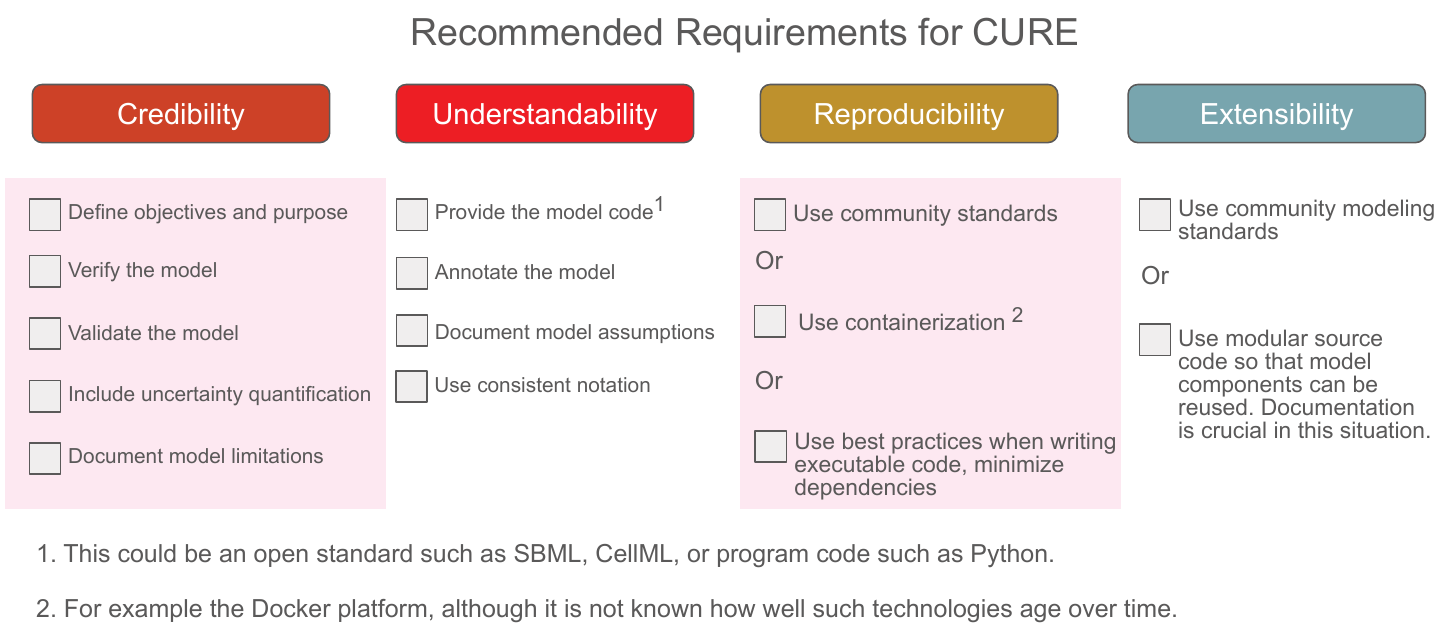}
\caption{Recommended or baseline requirements for {\tt CURE} that could be used for research-based models. The baseline requirements are highlighted in shaded pink boxes. A score of one out of ten can be given based on the number of criteria met. For example, the baseline requirement will yield a score of 6/10. Note that if publishing models via source code, it is essential to specify the version number of the software platform as well as version numbers of any dependency packages that were used. The use of containerization platforms such as docker can sometimes help in these situations.}
\label{fig:checklist}
\end{figure}

If the recommended requirements are still too onerous, it is possible to define a {\bf baseline requirement}. This term refers to the essential or foundational standards that are necessary for basic credibility, though they may not meet the full recommended requirements specified by {\tt CURE}. The baseline requirement is similar in intent to the recent report~\cite{national2023foundational} from the US National Academies, which emphasizes purpose, verification, validation, uncertainty quantification, and reproducibility, though their statement on reproducibility is vague. We include scope and model limitations into our baseline, which the National Academies documents do not explicitly mention, though it could be considered part of the statement of purpose. The baseline requirements are highlighted in Figure~\ref{fig:checklist}, and a more detailed summary is given in Table~\ref{tbl:summary}.

\begin{table}
\begin{tabular}{l}
\noindent\fbox{%
    \parbox{\textwidth}{%
    {\bf Credibility (Baseline requirement):}\\[3pt]
1. Clearly define the objectives and scope of the modeling study, including the biological question being addressed and the specific hypotheses to be tested.\\[-4pt]

2. Use consistent notation and terminology to ensure consistency and clarity in model descriptions. Where possible, follow common notations used in the community.\\[-4pt]        
        
3. Where possible, verify the model by checking the model with other simulators. Use model-checking tools to identify errors in the model~\citep{hellerstein2019recent}.\\[-4pt]

4. Validate the model against experimental data using accepted statistical procedures, such as cross-validation, to assess model accuracy and predictive power.\\[-4pt]   

5. Where possible, assess how sensitive the model is to uncertainty (UQ) in parameter estimates, model structure, inputs, and assumptions.\\[-4pt]

6. Clearly document the limitations of the model, including areas where assumptions may not hold or where uncertainties exist, to provide context for interpreting results and guiding future research.
    }%
}\\
\\
\vspace{5pt}

\noindent\fbox{%
    \parbox{\textwidth}{%
{\bf Understandability:}\\[3pt]
1. Provide a representation of the model that is both machine-readable and human-readable. Ideally, this should be in an open community standard and be an explicit representation of the model that is not intertwined with control logic, file input/output, and other implementation details.

2. Provide comprehensive documentation that explains the model structure, equations, and parameter values, e.g., by following the suggestions by~\citeauthor{Carey2020}~\citep{Carey2020}. When using open community standards such as SBML or CellML, submit the models to a recognized model repository. If the model is expressed in a programming language, deposit your model code at repositories such as GitHub, BioModels~\citep{malik2020biomodels}, or ModelDB~\citep{mcdougal2017twenty}.\\[-4pt] 

3. Where possible, annotate the model to clarify any ambiguous terminology. When using a programming language to express a model, use commenting to annotate the model. \\[-4pt]

4. Try to document all assumptions made during model development, including simplifications, approximations, and parameter values, to provide transparency and facilitate reproducibility.

5. Try to provide clear graphical illustration of the model. If the model is a biochemical network then use machine-readable formats such as SBGN~\cite{novere2009systems}, preferably using a community modeling standard such as SBML Layout~\cite{gauges2006model} and Render~\cite{bergmann2018sbml}.}
}\\
\vspace{6pt}

\noindent\fbox{%
    \parbox{\textwidth}{%
{\bf Reproducibility (Baseline requirement):}\\[3pt]
1. Follow established standards and guidelines for model development, such as the Systems Biology Markup Language (SBML) and Minimum Information Requested in the Annotation of Models (MIRIAM), to enhance interoperability and facilitate model sharing and exchange. If using executable code, make sure best practices for code development are used.\\[-4pt]

2. Embrace and promote open science practices by openly sharing publicly funded models, data, and code with the scientific community, promoting transparency, reproducibility, and collaboration.
    }%
}\\
\vspace{6pt}

\noindent\fbox{%
    \parbox{\textwidth}{%
{\bf Extensibility and Reuse:}\\[3pt]
1. Use open modeling standards where possible. If using executable code such as Python, separate the model code from the runtime code. In this form, the model code, in principle, can be reused with minimal effort by other Python users. However, this would require careful commenting on the components of the model. One approach is to provide the model as a software function that can be called by other code. Try to use open-source licensing so that there are no restrictions on the reuse of the research.\\[-4pt]

2. If a model is represented using a modeling format such as SBML, reuse should be much easier since the model is expressed in biological terms. If the model is annotated then automated systems can be devised to automate the merging and disassembly of models into individual parts of portions for reuse.
    }%
}
\end{tabular}
\caption{Summary of Recommended and Baseline Requirements.}
\label{tbl:summary}
\end{table}


\section{Conclusion}\label{sec13}

This paper introduces a set of guidelines for developing robust, credible biological models. These guidelines are meant to complement the FAIR guidelines for data. As with FAIR, we propose a four-word moniker, {\tt CURE}, that defines four core attributes of good modeling practice. These include credibility, understandability, reproducibility, and extensibility. 


The need for {\tt CURE} models is highlighted by recent interest in Biomedical Digital Twins, a technology that will rely on having robust models of biomedical systems~\citep{national2023foundational,laubenbacher2024toward,Bhagirath2024}, but it is clear that such guidelines would also benefit the wider biological modeling communities.

For credibility, we recommend the use of verification, validation, and uncertainty quantification; for understandability, we discuss the clarity of model descriptions and the importance of annotations; for reproducibility, our focus is standards and open science practices; and for extensibility, we emphasize open standards and the use of modular models. We outline recommended requirements for each guideline and propose a baseline level below the recommended requirements that is largely in alignment with the National Academies report~\citep{national2023foundational}.



While all of the {\tt CURE} principles are important, we wish to highlight credibility, the first principle. Credibility is the degree to which a model can be trusted when applied to a given problem. In a biomedical application where concern is with patients and their well-being, the trustworthiness of a predictive model is paramount. Interestingly, the FDA recently published~\citep{food2021assessing} a guidance document on assessing the credibility of computation models but applied only to medical devices. Many of their recommendations relate to model verification and validation but also contained aspects related to model plausibility, which considers the plausibility of the governing equations, assumptions, and model parameters. The FDA document also emphasizes UQ for estimating uncertainty in the model outputs. These are considered foundational for credible modeling. However, these have not yet infiltrated the biomedical modeling community significantly. The recent National Academies~\citep{national2023foundational} report on Digital Twins emphasizes the same points. The report also stresses the critical need for data and modeling standards to enable interoperability and reuse.  As an initial effort to support VVUQ, the BioSimulations resource~\citep{shaikh2022biosimulators} provides a verification service where a given model can be run against multiple independent simulators to help verify the simulation engines. 

Finally, automation is key to making the {\tt CURE} guidelines workable and practical to reduce the burden on practitioners and accelerate widespread adoption. 

\bmhead{Acknowledgments}

This work was supported by NIH Biomedical Imaging and Bioengineering award P41 EB023912 through HMS at the Center for Reproducible Biomedical Modeling (\url{https://reproduciblebiomodels.org/}). The content expressed here is solely the responsibility of the authors and does not necessarily represent the official views of the National Institutes of Health, or the University of Washington. HMS wishes to thank Eric Johnson Chavarria for suggesting the CURE acronym at the 2023 IMAG meeting in Bethesda, MD. HMS also wishes to thank Hunter Robbins for assistance in collating the author names and addresses. 

T.E.G. was supported by a Royal Society University Research Fellowship (URF\textbackslash R\textbackslash 221008) and the UKRI-BBSRC Engineering Biology Mission Award CYBER (BB/Y007638/1).

S.F. is supported by the Predictive Phenomics Initiative under the Laboratory Directed Research and Development Program at Pacific Northwest National Laboratory, operated by Battelle for the U.S. Department of Energy under Contract No. DE-AC05-76RL01830.
J.F. was supported by DARPA through the Automating Scientific Knowledge Extraction and Modeling (ASKEM) program, Agreement No. HR0011262087;  NSF awards IIS-2106888 and CMMI-2146306. The views, opinions, and findings expressed are those of the authors and should not be interpreted as representing the official views or policies of the Department of Defense, the U.S. Government, or NSF.

H.M.S. acknowledges research reported in this publication was supported by NIBIB of the National Institutes of Health under award number NIH grant number P41EB023912.

R.L. acknowledges funding from the following awards: NIH 1 R01 HL169974-01, U.S. DoD DARPA HR00112220038, NIH 1 R011AI135128-01, NIH 1 R01 HL169974-01.

R.V. acknowledges funding from the following awards: National Institute on Alcohol Abuse and Alcoholism R01 AA018873, National Heart, Lung, and Blood Institute R01 HL161696. The funding sponsors had no role in the design of the study; in the collection, analyses, or interpretation of data; in the writing of the manuscript, and in the decision to publish the results.

NR was funded by Deutsche Forschungsgemeinschaft (DFG, German Research Foundation) under Germany's Excellence Strategy - EXC 2075 – 390740016.

G.D.B acknowledges work was supported by NRNB (U.S. National Institutes of Health, National Center for Research Resources grant number P41 GM103504).

J.H.G acknowledges research reported in this publication was supported by NIBIB of the National Institutes of Health under award number NIH grant number P41EB023912.

L.M.L acknowledges work was supported by NIH grant R24 GM137787 from the National Institute of General Medical Sciences.

J.L.S acknowledges funding from the following award: DST/NRF SARCHI-82813.

D.v.N. acknowledges funding from the following award: DST/NRF SRUG2204173612.

P.M. acknowledges work was supported by NIH grant R24 GM137787 from the National Institute of General Medical Sciences.

F.F. acknowledges work was supported by the Francis Crick Institute, which receives its core funding from Cancer Research UK (CC2242), the UK Medical Research Council (CC2242), and the Wellcome Trust (CC2242).

J.R.F acknowledges support from NIH grants P41GM10371 and R01GM115805.

T.J.S. acknowledges funding from NSF grant 2000281.

A.D. acknowledges support by the German Center for Infection Research (DZIF), grant № 8020708703.

J.M.R. acknowledges funding from the following award: NRF grant number SRUG2204295377.

\section*{Contributions}
HMS conceived the study. HMS, EG, JHG, PH, BRJ, EM, GDB, PMB, AD, JF, TEG, CJM, RV, JH, CFL, WWL, AM and PM wrote and edited the manuscript. The remaining authors read and approved the manuscript content.

\section*{Ethics declarations}

Competing interests

The authors declare no competing interests.


\bibliography{sn-bibliography}

\end{document}